\documentclass[epj,twocolumn]{webofc} 

\usepackage[varg]{txfonts}   
\usepackage{widetext}
\usepackage{bm}

\newcommand{\sig}{\sigma}

\newcommand{\sT}{\scriptstyle T}
\newcommand{\nn}{\nonumber}
\newcommand{\Pslash}{\kern 0.2 em P \kern -0.56 em \raisebox{0.3ex}{/}}
\newcommand{\Rslash}{\kern 0.2 em R \kern -0.56 em \raisebox{0.3ex}{/}}
\newcommand{\kslash}{\kern 0.2 em k \kern -0.45 em /}
\newcommand{\pslash}{\kern 0.2 em p \kern -0.45 em /}
\newcommand{\Sslash}{\kern 0.2 em S \kern -0.45 em /}
\newcommand{\nslash}{\kern 0.2 em n \kern -0.45 em /}
\newcommand{\thslash}{\kern 0.2 em \theta \kern -0.54 em /}

\newcommand{\p}{\perp}

\newcommand{\hermes}{\textsc{Hermes }}

\newcommand{\belle}{\textsc{Belle }}
\newcommand{\babar}{\textsc{BaBar }}

\woctitle{TRANSVERSITY 2014} 
\begin{document}
\title{Phenomenology from SIDIS and $e^+e^-$ multiplicities}
%
%
\subtitle{Multiplicities and phenomenology - part I}

\author{Alessandro Bacchetta\inst{1,2}\fnsep\thanks{\email{alessandro.bacchetta@unipv.it}} \and
        Miguel G. Echevarria\inst{3,4}\fnsep\thanks{\email{miguelge@nikhef.nl}} \and
        Marco Radici\inst{1}\fnsep\thanks{\email{marco.radici@pv.infn.it}} \and
        Andrea Signori\inst{3,4}\fnsep\thanks{\email{asignori@nikhef.nl}}
}

\institute{INFN Sezione di Pavia, via Bassi 6, I-27100 Pavia, Italy
\and
Dipartimento di Fisica, Universit\`a di Pavia, via Bassi 6, I-27100 Pavia, Italy 
\and
Department of Physics and Astronomy, VU University Amsterdam, De Boelelaan 1081, NL-1081 HV Amsterdam, the Netherlands
\and
Nikhef, Science Park 105, NL-1098 XG Amsterdam, the Netherlands
}

\abstract{
This study is part of a project to investigate the transverse momentum dependence in parton distribution and fragmentation functions, analyzing (semi-)inclusive high-energy processes within a proper QCD framework.
We calculate the transverse-momentum-dependent (TMD) multiplicities for $e^+e^-$ annihilation into two hadrons (considering different combinations of pions and kaons) aiming to investigate the impact of intrinsic and radiative partonic transverse momentum and their mixing with flavor.
Different descriptions of the non-perturbative evolution kernel (see, e.g., Refs.~\cite{Nadolsky:2000ky,Landry:2002ix,Konychev:2005iy,Aidala:2014hva,D'Alesio:2014vja}) are available on the market and there are 200 sets of flavor configurations for the unpolarized TMD fragmentation functions (FFs) resulting from a Monte Carlo fit of Semi-Inclusive Deep-Inelastic Scattering (SIDIS) data at \hermes (see Ref.~\cite{Signori:2013mda}). We build our predictions of $e^+e^-$ multiplicities relying on this rich phenomenology. The comparison of these calculations with future experimental data (from \belle and \babar collaborations) will shed light on non-perturbative aspects of hadron structure, opening important insights into the physics of spin, flavor and momentum structure of hadrons.
}


\maketitle

\section{Introduction}
\label{s:intro}
The 	QCD description of electron-positron annihilation into hadrons with observed transverse momenta in the final states requires the usage of TMD FFs. These are distributions in 3-dimensional momentum space describing the probability for elementary quarks and gluons to generate hadrons, composite states bounded by the color force. 
Here we calculate multiplicities for annihilation into two hadrons, focusing on the input brought by non-perturbative QCD. 
The contribution coming from gluons radiated with low transverse momentum by the fragmenting quark must be phenomenologically parametrized and different choices drive to different shapes in the multiplicity. 
The second source of non-perturbative information that we take into account is the flavor structure of TMD FFs. Previous analyses (see Ref.~\cite{Signori:2013mda,Signori:2013gra,Signori:2014kda}) revealed that the the TMD part of FFs depends on the quark flavor, confirming the physically intuitive picture of quarks fragmenting into different hadrons with different probability amplitudes.
Here we investigate to which extent the annihilation rate is modified according to different flavor configurations for the intrinsic transverse momentum of quarks. 

\section{SIDIS multiplicities and partonic flavor structure}
\label{s:sidis}
In Refs.~\cite{Signori:2013mda,Signori:2013gra,Signori:2014kda} a phenomenological investigation of unpolarized SIDIS multiplicities is presented, focusing on the flavor dependence of the intrinsic partonic transverse momentum in TMD parton distribution functions (PDFs) and FFs. There are convincing hints on the flavor dependence of unpolarized TMD FFs, supporting the physically intuitive picture of different fragmentation probabilities for different flavor configurations. The results concerning distribution functions are weaker, but phenomenology is pointing towards different distributions at least for valence and sea quarks. 
\begin{figure}
\centering
\includegraphics[width=8cm]{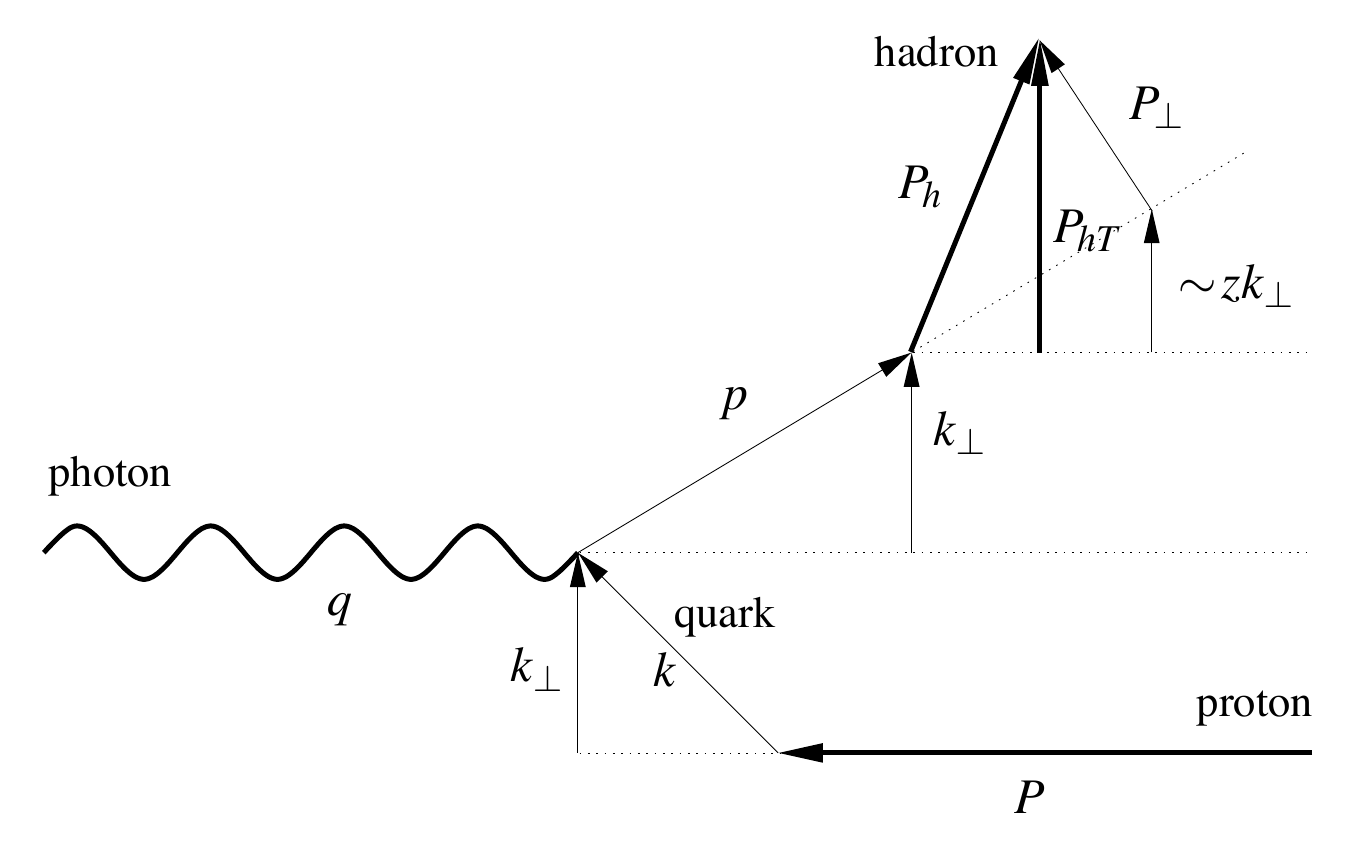}
\caption{Definition of transverse momenta involved in unpolarized SIDIS.}
\label{f:transmom_SIDIS}
\end{figure}
\begin{figure}
\centering
\includegraphics[width=8cm]{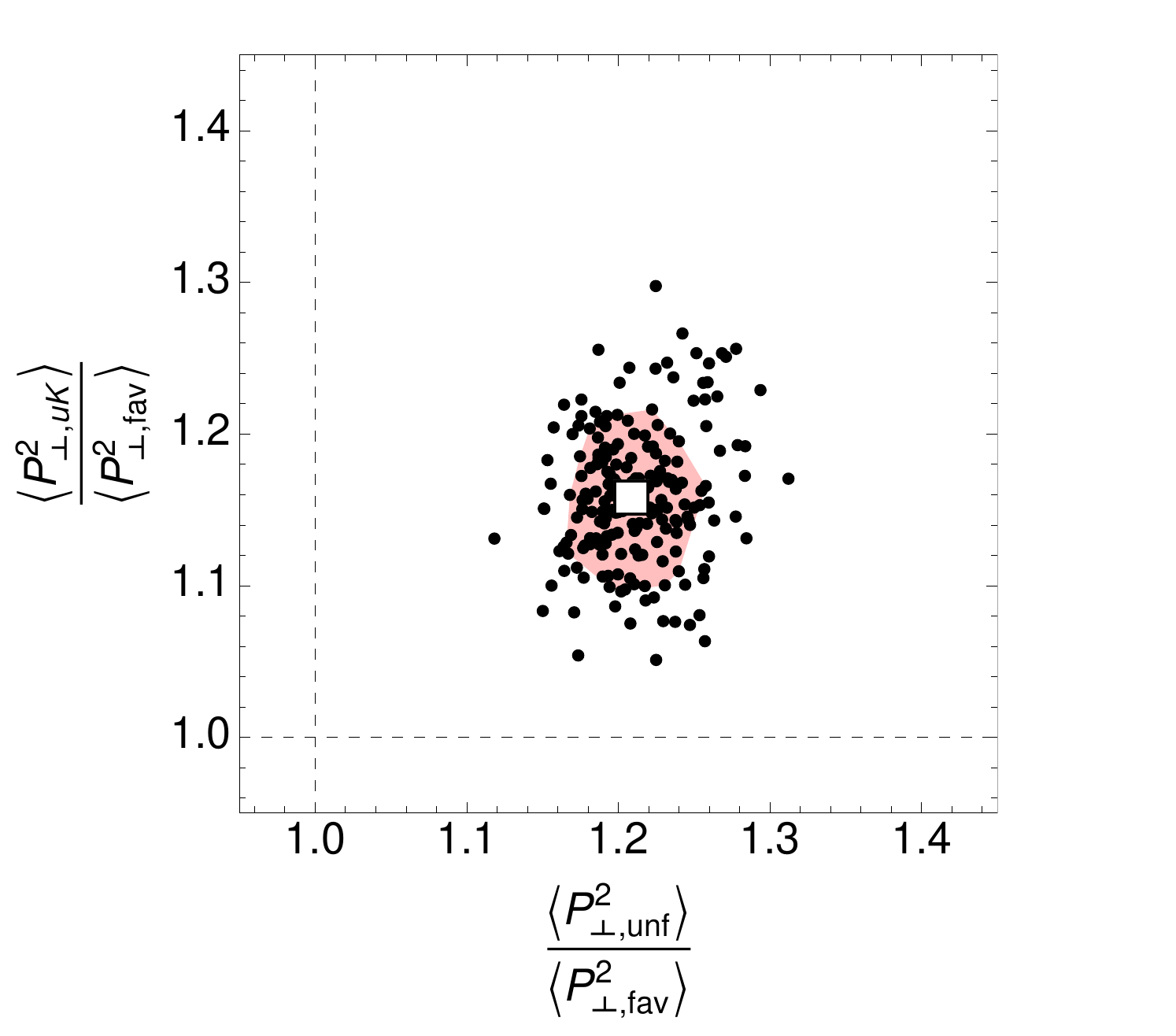}
\caption{Kinematic-independent distribution of average squared transverse momenta in TMD FFs: $\langle \bm{P}_{\p, {\rm unf}}^2 \rangle / \langle \bm{P}_{\p, {\rm fav}}^2 \rangle$ vs. $\langle \bm{P}_{\p, u K}^2 \rangle / \langle \bm{P}_{\p, {\rm fav}}^2 \rangle$.}
\label{f:fldep_TMDFFs}
\end{figure}
In this approach unpolarized TMD PDFs and FFs are parametrized as the product of the collinear distribution and a Gaussian function with both flavor and kinematic-dependent variance. Transverse momenta are defined as in Fig.~\ref{f:transmom_SIDIS}. TMD PDFs read
\begin{equation}
f_1^a(x,\bm{k}_{\p}^2;Q^2) = f_1^a(x;Q^2)\ \exp{ \bigg\{ - \frac{\bm{k}_{\p}^2}{\big \langle \bm{k}_{\p,a}^2 \big \rangle(x)} \bigg\} }\ ,
\end{equation}
whereas TMD FFs are
\begin{equation}
D_1^{a \to h}(z,\bm{P}_{\p}^2;Q^2) = D_1^{a \to h}(z;Q^2)\ \exp{ \bigg\{ - \frac{\bm{P}_{\p}^2}{\big \langle \bm{P}_{\p,a \to h}^2 \big \rangle(z)} \bigg\} }\ .
\end{equation}
The kinematic $x$ and $z$ dependence of the variances is fully described in Ref.~\cite{Signori:2013mda} (and partially in Sec.~\ref{s:predictions}).
The flavor dependence in TMD PDFs consists in considering three different widths in the Gaussian distributions, one for up-valence quarks, one for down-valence quarks and one for sea quarks:
\begin{gather}
\langle \bm{k}_{\p,u_v}^2 \rangle \ , \langle \bm{k}_{\p,d_v}^2 \rangle \ , \langle \bm{k}_{\p,\text{sea}}^2 \rangle \ .
\end{gather}
Concerning TMD FFs, instead, three distinct favored process and one class of unfavored processes have been distinguished, assuming charge conjugation and isospin symmetry. This results in four different Gaussian widths:
\begin{gather}
\big\langle \bm{P}^2_{\p,u \to \pi^+} \big\rangle = 
\big \langle \bm{P}^2_{\p,\bar{d} \to \pi^+} \big \rangle = 
\big \langle \bm{P}^2_{\p,\bar{u} \to \pi^-} \big \rangle =  
\big \langle \bm{P}^2_{\p,d \to \pi^-}\big \rangle \equiv \big \langle \bm{P}^2_{\p, {\rm fav}} \big \rangle \nn \, ,  
\label{e:favored}  
\\
\big \langle \bm{P}^2_{\p,u \to K^+} \big \rangle =  
\big \langle \bm{P}^2_{\p,\bar{u} \to K^-} \big \rangle \equiv \big \langle \bm{P}^2_{\p, {uK}} \big \rangle \nn \, ,
\label{e:uK}  
\\
\big \langle \bm{P}^2_{\p,\bar{s} \to K^+} \big \rangle = 
\big \langle \bm{P}^2_{\p,s \to K^-}\big \rangle \equiv \big \langle \bm{P}^2_{\p, {sK}} \big \rangle \nn \,  ,
\label{e:sK}  
\\
\big \langle \bm{P}^2_{\p,\text{all others}} \big \rangle 
\equiv  \big \langle \bm{P}^2_{\p, {\rm unf}} \big \rangle \, .
\label{e:unfavored} 
\end{gather} 
The fit is performed on SIDIS data sets from \hermes, considering proton and deuteron targets and pions and kaons in the final state. 
Since the covered $Q^2$ range is narrow, the analysis has been carried out at a fixed $Q^2 = 2.4\ \text{GeV}^2$ neglecting any effect from QCD evolution.
Despite this limitation, the available phenomenology is rich, because of the multidimensional binning in $x$, $z$ and $Q^2$.
The fit relies on ${\cal M}=200$ Monte Carlo {\em replicas} of the original data set, in order to get ${\cal M}$ best values for each fit parameter. This is a powerful procedure, because it allows to calculate statistical distributions for physical observables.
The Monte Carlo fit indicates that, on average, the Gaussian distribution of sea quarks is 20\% wider than the one for up-valence quarks, which, in turn, is 20\% wider than the distribution for down-valence quarks. Moreover, the flavor independent configuration lies at the boundary of the 68\% confidence region (see Ref.~\cite{Signori:2013mda}), so it is not ruled out by statistics.
Results concerning TMD FFs are described in Fig.~\ref{f:fldep_TMDFFs}, comparing the ratio $\langle \bm{P}_{\p, {\rm unf}}^2 \rangle / \langle \bm{P}_{\p, {\rm fav}}^2 \rangle$ vs. $\langle \bm{P}_{\p, u K}^2 \rangle / \langle \bm{P}_{\p, {\rm fav}}^2 \rangle$.
On average, the width of unfavored and $u \to K$ fragmentations are about 20\% larger than the width of favored ones.
All points are concentrated in the upper right quadrant: we have the clear outcome that $\langle \bm{P}_{\p, {\rm fav}}^2 \rangle < \langle \bm{P}_{\p, {\rm unf}}^2 \rangle \sim \langle \bm{P}_{\p, u K}^2 \rangle$ and that the flavor-independent configuration (the crossing point of the dashed lines) is well outside the 68\% confidence region, meaning that the flavor dependence is statistically much more evident than in TMD PDFs.
It is important to stress that from this Monte Carlo fit of SIDIS multiplicities, there are ${\cal M}=200$ configurations available for TMD FFs, all equivalent from the $\chi^2$ point of view, but different from the perspective of flavor dependence: in each of them, the ratios of transverse momenta assume different values, so they represents different flavor configurations.
More details about the theoretical framework, the phenomenological assumptions and the data analysis are available in the Refs.~\cite{Signori:2013gra,Signori:2013mda,Signori:2014kda}.

\section{$e^+e^-$ multiplicities}
\label{s:e+e-}

In this contribution we want to investigate the multiplicity $M[h_1,h_2]$ for $e^+e^-$ annihilation into two hadrons.
The definition of this observable, as described in Sec.~\ref{ss:twohad_mult}, involves the expression of cross sections for production of one and two hadrons:
\begin{gather}
e^+e^- \rightarrow h_1\ \text{jet}\ X 
\label{e:onehad}
\\
e^+e^- \rightarrow h_1\ h_2\ X 
\label{e:twohad}
\end{gather} 
They key mathematical objects involved in the analysis are TMD FFs, in particular their flavor structure and QCD evolution.
In order to calculate cross sections and multiplicities for $e^+e^-$ annihilation into hadrons at $Q^2=100$ GeV$^2$ (typical value at the \belle experiment) we use resummation techniques (see Refs.~\cite{Collins:2011zzd,Echevarria:2014rua}) to evolve the low-energy distributions extracted from SIDIS data at $Q^2=2.4$ GeV$^2$.
Evolution of distribution functions is performed in impact parameter space. The evolution operator acts on $D_1^{a \to h}(z,\bm{b}_{\sT};Q^2)$, defined as the Fourier-conjugated of the TMD FF $D_1^{a \to h}$ as a function of the {\em partonic} transverse momentum $\bm{k}_T$.
According to the definition of $D_1^{a \to h}(z,\bm{P}_{\p};Q^2)$ and its normalization given in Ref.~\cite{Signori:2013mda} in terms of the hadronic transverse momentum $\bm{P}_{\p}$, the expressions for the upolarized TMD FF and its Fourier transform as a function of partonic transverse momentum are:
\begin{gather}
D_1^{a \to h}(z,\bm{k}_{\sT}^2;Q_0^2) = D_1^{a \to h}(z;Q_0^2)\ 
\frac{ \exp \bigg\{-\frac{\bm{k}_{\sT}^2}{\big \langle \bm{k}^2_{\sT, a \to h} \big \rangle } \bigg\} }{\pi\ z^2\ \langle k_{T, a \to h}^2 \rangle} \ , 
\label{e:TMDFF_part}
\\
D_1^{a \to h}(z,\bm{b}_T^2;Q_0^2) = \frac{D_1^{a \to h}(z;Q_0^2)}{z^2}\ \exp \bigg\{ -\frac{\bm{b}_T^2 \ \big \langle \bm{P}^2_{\p, a \to h} \big \rangle}{4 z^2} \bigg\} \ .
\label{e:TPDFF_part}
\end{gather}
Phenomenological predictions for the specific process in Eq.~\eqref{e:onehad} are presented in Ref.~\cite{Bacchetta:proc_QCDEvo14}.

\subsection{Multiplicities for production of two hadrons}
\label{ss:twohad_mult}

We define multiplicities for $e^+e^-$ annihilation into two hadrons exactly as in the SIDIS case, namely as the ratio between the cross section for the process in Eq.~\eqref{e:twohad} differential with respect to the transverse momentum of one hadron and the collinear ``one-particle more inclusive'' cross section, the one related to Eq.~\eqref{e:onehad} integrated over transverse momentum:
\begin{equation}
M[h_1,h_2] = \frac{ \frac{d\sigma^{h_1 h_2}}{dz_1 dz_2 d\bm{q}^2_{\sT} dy}(e^+e^- \rightarrow h_1\ h_2\ X) }{ \frac{d\sigma^{h_1}}{dz_1 dy}(e^+e^- \rightarrow h_1\ \text{jet}\ X) } \ .
\label{e:mult_2had}
\end{equation}
$z_1$ and $z_2$ are the light-cone momentum fractions of the two produced hadrons, $\bm{q}_{\sT}$ is the transverse momentum of the virtual photon emerging from $e^+e^-$ annihilation and $y$ is the rapidity variable. 
A complete description of notations and conventions is available in Ref.~\cite{Bacchetta:in_prep}.
The experimentally accessible transverse momentum is defined to be the momentum of $h_1$ transverse with respect to the momentum of $h_2$. As in Ref.~\cite{Boer:1997mf}, it is defined as:
\begin{equation}
\bm{P}_{\p, 1} = - z_1 \bm{q}_{\sT} \ .
\label{e:exp_tm_h1}
\end{equation}
As described in Ref.~\cite{Bacchetta:in_prep}, the cross section for the production of two hadrons (see Eq.~\eqref{e:twohad}) differential with respect to the square modulus of the transverse momentum of the photon is:
\begin{widetext}
\begin{eqnarray}
\frac{d\sig^{h_1 h_2}}{dz_1\, dz_2\, d\bm{q}^2_{\sT}\, dy} 
 	 =  \frac{6 \pi \alpha^2}{Q^2} A\ {\cal H}\ z_1^2 z_2^2 \sum_q e_q^2 \int_0^\infty d b_{\sT} b_{\sT} \, J_0 (q_{\sT} b_{\sT}) \,\bigg[ D_1^{q \to h_1} (z_1, \bm{b}^2_{\sT};Q^2) \, D_1^{\overline{q} \to h_2}(z_2, \bm{b}^2_{\sT}; Q^2) + (q \leftrightarrow \overline{q}) \bigg] + Y\ ,
\label{e:xsec_bT_h1h2X}
\end{eqnarray}
\end{widetext}
where $A=A(y)$ is a function of the rapidity 
and ${\cal H} = {\cal H}(Q)$ is a hard coefficient, function of the energy scale (see Ref.~\citep{Bacchetta:in_prep} for further details). $Y = Y(\bm{q}^2_{\sT},Q^2)$ is the term correcting the factorized formula for the kinematic region where $\bm{q}_{\sT}^2 \sim Q^2$.
Relying on the same assumptions, the cross section for production of one hadron (see Eq.~\eqref{e:onehad}) integrated over its transverse momentum is:
\begin{eqnarray}
\frac{d\sig^{h_1}}{dz_1 dy} = 
 	\frac{12 \pi \alpha^2}{Q^2} A \sum_q e_q^2\ D_1^{q \to h_1} (z_1;Q^2) \ .
\label{e:xsec_h1jetX}
\end{eqnarray}
In this study we will make predictions for pion multiplicities $M[\pi^+,\pi^-]$, built from Eqs.~\eqref{e:xsec_bT_h1h2X} and~\eqref{e:xsec_h1jetX} with $\{h_1,h_2\}=\{\pi^+,\pi^-\}$ and kaons multiplicities $M[K^+,K^-]$, obtained evaluating the same equations with $\{h_1,h_2\}=\{K^+,K^-\}$.
Mixed multiplicities (like $M[\pi^+,K^-]$ and the like) will be left for future investigations.
The summation over flavors is limited to up, down and strange quarks (the ones included in the analysis of SIDIS data in Ref.~\cite{Signori:2013mda}, for which we can provide a TMD FF).

\subsection{QCD evolution of TMD FFs}
\label{ss:evo_FFs}

From SIDIS data we gained a working flavor-dependent Gaussian parametrization of TMD FFs at an initial scale $Q_0$. Using QCD evolution we can calculate the fragmentation function $D_1^{q \to h} (z;Q^2)$ and $D_1^{q \to h} (z, \bm{b}_{\sT}; Q^2)$ at some final scale $Q$, thus making predictions for the transverse momentum dependence of processes in Eqs.~\eqref{e:onehad} and~\eqref{e:twohad}. 
The collinear FF in Eq.~\eqref{e:xsec_h1jetX} is evolved through the DGLAP evolution, acting on the renormalization scale $\mu$, which we set equal to $Q$.
TMD distributions, instead, generally depend on two scales, $\zeta$ an $\mu$, and they satisfy evolution equations with respect to both of them (see Refs.~\cite{Collins:2011zzd,Echevarria:2014rua}). 
The evolution with respect to $\zeta$ is determined by a process-independent soft factor, whereas the evolution in $\mu$ is determined by renormalization group equations. Here we follow the approach of Ref.~\cite{Collins:2011zzd} and use the so-called $\bm{b}_{\sT}^*$ prescription to separate perturbative and non-perturbative regions. $\bm{b}_{\sT}^*$ is defined as
\begin{equation}
\bm{b}_{\sT}^* = \frac{\bm{b}_{\sT}}{\sqrt{1 + \frac{\bm{b}_{\sT}^2}{b_{\text{max}}^2}}} \ .
\end{equation}
The parameter $b_{\text{max}}$ represents the value where we stop trusting perturbative QCD (pQCD). Despite knowing the full operator structure, for $\bm{b}_{\sT} > b_{\text{max}}$ a model is required to work out phenomenological calculations.
For sake of simplicity we make the following choices for the initial and final scales:
\begin{equation}
\mu_i^2 = \zeta_i = \mu_b^2\ ,\ \ \ \ \ 
\mu^2 = \zeta = Q^2 \ ,
\label{e:choice_scales}
\end{equation}
where $\mu_b = 2 e^{-\gamma_E}/b^*$. Considering the distribution at $Q$ as the action of an evolution operator on an input distribution at $Q_0$ we get:
\begin{widetext}
\begin{align}
D_1^{q \to h}(z,\bm{b}^2_T;Q^2)
=&
\exp\ \bigg\{ 
       -\int_{\mu_b}^{Q} \frac{d\bar\mu}{\bar\mu} 
      \bigg(\Gamma_{\rm cusp}\,\ln\frac{Q^2}{\bar\mu^2}+\gamma^V \bigg) 
      \bigg\} \ \ 
\bigg( \frac{Q^2}{\mu_b^2} \bigg)^{-D(\bm{b}_T^*;\mu_b)-\frac{1}{4} g_2 \bm{b}_T^2}
\nn\\
&\times
\sum_{j=q,\bar q, g}
\int_{z}^{1}\frac{dx}{x}
{\tilde I}_{q\leftarrow j}(z/x,\bm{b}_T^*;\mu_b) \,
D_1^{j \to h}(x;\mu_b)
\bigg(\frac{\mu_b^2}{Q_0^2}\bigg)^{-\frac{1}{4} g_2 \bm{b}_T^2}
e^{-\frac{1}{2} g_1 \bm{b}_T^2} \ ,
\label{e:evo_TMDFF}
\end{align}
\end{widetext}
where the first line represents the evolution operator and the second one is the input TMD distribution.
Here the large-$\bm{b}_T$ region is inspired to the model in Ref.~\cite{Nadolsky:2000ky}, to the BLNY model (see Ref.~\cite{Landry:2002ix}) and to Ref.~\cite{Signori:2013mda}.
In the present context the parameter $g_1$ is related to the flavor and kinematic dependent Gaussian widths of Ref.~\cite{Signori:2013mda}:
\begin{equation}
g_1 \equiv \frac{\big \langle \bm{P}^2_{\p,a \to h} \big \rangle (z)}{2} \ .
\label{e:g1_tm}
\end{equation}
Other parametrizations are available for the large-$\bm{b}_T$ region (see, e.g., Refs.~\cite{Konychev:2005iy,Aidala:2014hva,D'Alesio:2014vja}) and multiple parameter sets could describe presently available data. Considering our choice of functional form, we do not know which values of the parameters $\{ b_{\text{max}}, g_2 \}$ are the best ones in order to reproduce the transverse momentum spectrum of $e^+e^-$ annihilation into hadrons. 
The same holds for the flavor-dependent widths of the Gaussian FFs. From the SIDIS point of view there are 200 equivalent sets of values, but we do not know which 
ones work best considering processes in Eqs.~\eqref{e:xsec_bT_h1h2X} and~\eqref{e:xsec_h1jetX}.
This study is exactly aimed at underlying the sensitivity of $e^+e^-$ multiplicities to the non-perturbative parameters (concerning flavor structure and evolution) and our predictions are collected in Sec.~\ref{s:predictions}.

\section{Predictions for $e^+e^- \rightarrow h_1\ h_2\ X$}
\label{s:predictions}

The parameters $b_{\text{max}}$ and $g_2$ are anti-correlated. This is because the first one selects the $\bm{b}_{\sT}$ value where we do not trust pQCD any more and the second one shapes the effects of gluon radiation for $\bm{b}_{\sT} > b_{\text{max}}$. So, lowering $b_{\text{max}}$ results in a larger $\bm{b}_{\sT}$-range where the evolution needs to be parametrized and, eventually, in a higher value for the $g_2$ parameter.
In Tab.~\ref{t:g2bmax_values} we summarize the three different configurations of values for $b_{\text{max}}$ and $g_2$ explored in this study.
\begin{table}
\centering
\begin{tabular}{|l|l|l|l|}
\hline
configuration  &  $b_{\text{max}}$ [GeV$^{-1}$]  &  $g_2$  &  Ref.  \\
\hline
A   &   0.5   &   0.68   &  \cite{Landry:2002ix}  \\
B   &   1.0   &   0.41   &  -  \\
C   &   1.5   &   0.18   &  \cite{Konychev:2005iy}  \\
\hline
\end{tabular}
\caption{The three configurations for $b_{\text{max}}$ and $g_2$ explored in this study and their source (reference).}
\label{t:g2bmax_values}
\end{table}
We recall that the flavor dependence of unpolarized TMD FFs is encoded in four different Gaussian widths (and fragmentation processes):
\begin{equation}
\big \langle \bm{P}^2_{\p, {\rm fav}} \big \rangle \ ,
\ \ \ 
\big \langle \bm{P}^2_{\p, {uK}} \big \rangle \ ,
\ \ \ 
\big \langle \bm{P}^2_{\p, {sK}} \big \rangle \ ,
\ \ \ 
\big \langle \bm{P}^2_{\p, {\rm unf}} \big \rangle \ .
\label{e:frag_proc}
\end{equation}
Each width is also $z$-dependent:
\begin{align}  
\label{e:z_dep}
&\big \langle \bm{P}_{\p,a \to h}^2 \big \rangle (z) = \big \langle
\hat{\bm{P}}_{\p,a \to h}^2 \big \rangle 
\frac{ (z^{\beta} + \delta)\ (1-z)^{\gamma} }{ (\hat{z}^{\beta} + \delta)\
  (1-\hat{z})^{\gamma} } \ , 
\\ \nn
&\text{where }
\big\langle \hat{\bm{P}}_{\p,a \to h}^2 \big\rangle \equiv \big \langle
\bm{P}_{\p,a \to h}^2
\big \rangle
(\hat{z})
\\ \nn
&\text{ and } \hat{z}=0.5.
\end{align}  
The kinematic parameters $\beta$, $\gamma$ and $\delta$ are flavor-independent, contrary to the normalizations.
For each parameter there are 200 sets of values available (see Ref.~\cite{Signori:2013mda}) and in this study we exploit the first 100.


\subsection{Impact of evolution}
\label{ss:evolution}

\begin{figure}
\centering
\includegraphics[width=8.5cm]{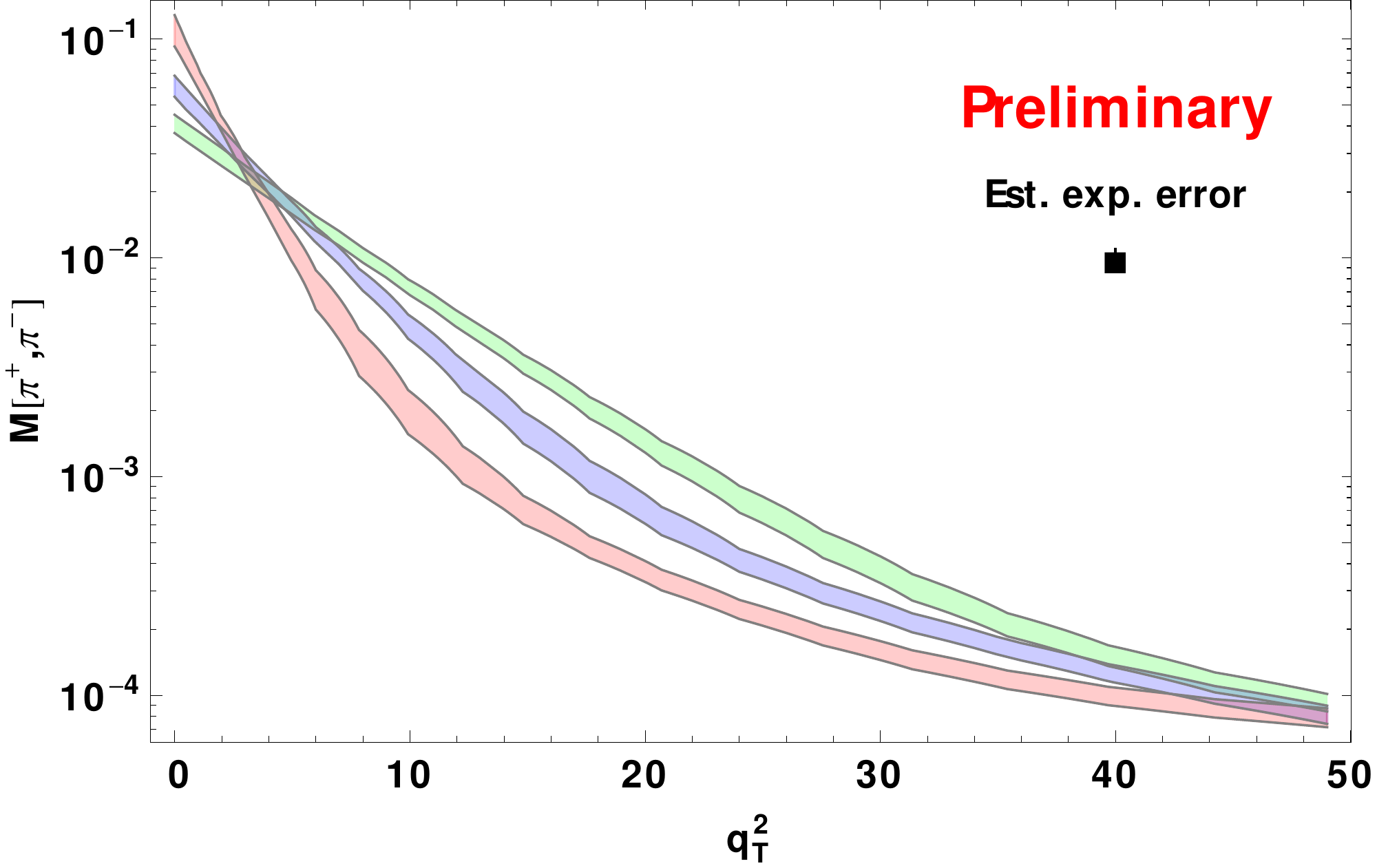}
\caption{Logarithm of the pion multiplicity $M[\pi^+,\pi^-]$ with $z_1 = z_2 = 0.7$ as a function of $\bm{q}_{\sT}^2$. The correspondence between colors and configurations is $\{$green,blue,red$\}=\{$A,B,C$\}$. An hypothetical experimental point with uncertainty of 10\% is shown.}
\label{f:impact_evo}
\end{figure}

Fig.~\ref{f:impact_evo} shows the impact on pion multiplicity $M[\pi^+,\pi^-]$ of the non-perturbative parameters related to evolution, for $z_1 = z_2 = 0.7$.
In this case the maximum $\bm{q}^2_{\sT}$ value (50 GeV$^2$) corresponds to $\bm{P}_{\p} = 5$ GeV for the detected hadron $h_1$ (see Eq.~\eqref{e:exp_tm_h1}).
Configuration A is plotted in green, B in blue and C in red.
For each configuration there is a band and not a single plot, because we exploited 100 out of the 200 values of the Gaussian widths in TMD FFs.
This demonstrates that, considering the evolution framework in Ref.~\cite{Collins:2011zzd} with $Q_i^2=2.4$ GeV$^2$ and $Q_f^2=100$ GeV$^2$, the pion multiplicity is sensitive to the choice of the non-perturbative evolution kernel.
Estimating an experimental error of 10\% of the multiplicity value (in Figs.~\ref{f:impact_evo},~\ref{f:impact_fldep},~\ref{f:impact_kin} the error is an overestimation of the experimental uncertainty affecting collinear multiplicities - see Ref.~\cite{Leitgab:2013qh}), we see that experimental data will be able to discriminate among the configurations A, B and C.
The effect is less evident for lower $z_1$ and $z_2$ values, because of the $z^2$ factor in the exponent in Eq.~\eqref{e:TPDFF_part}.
The effect is very similar considering kaon multiplicities $M[K^+,K^-]$.

\subsection{Impact of flavor dependence}
\label{ss:flavor}

\begin{figure}
\centering
\includegraphics[width=8.5cm]{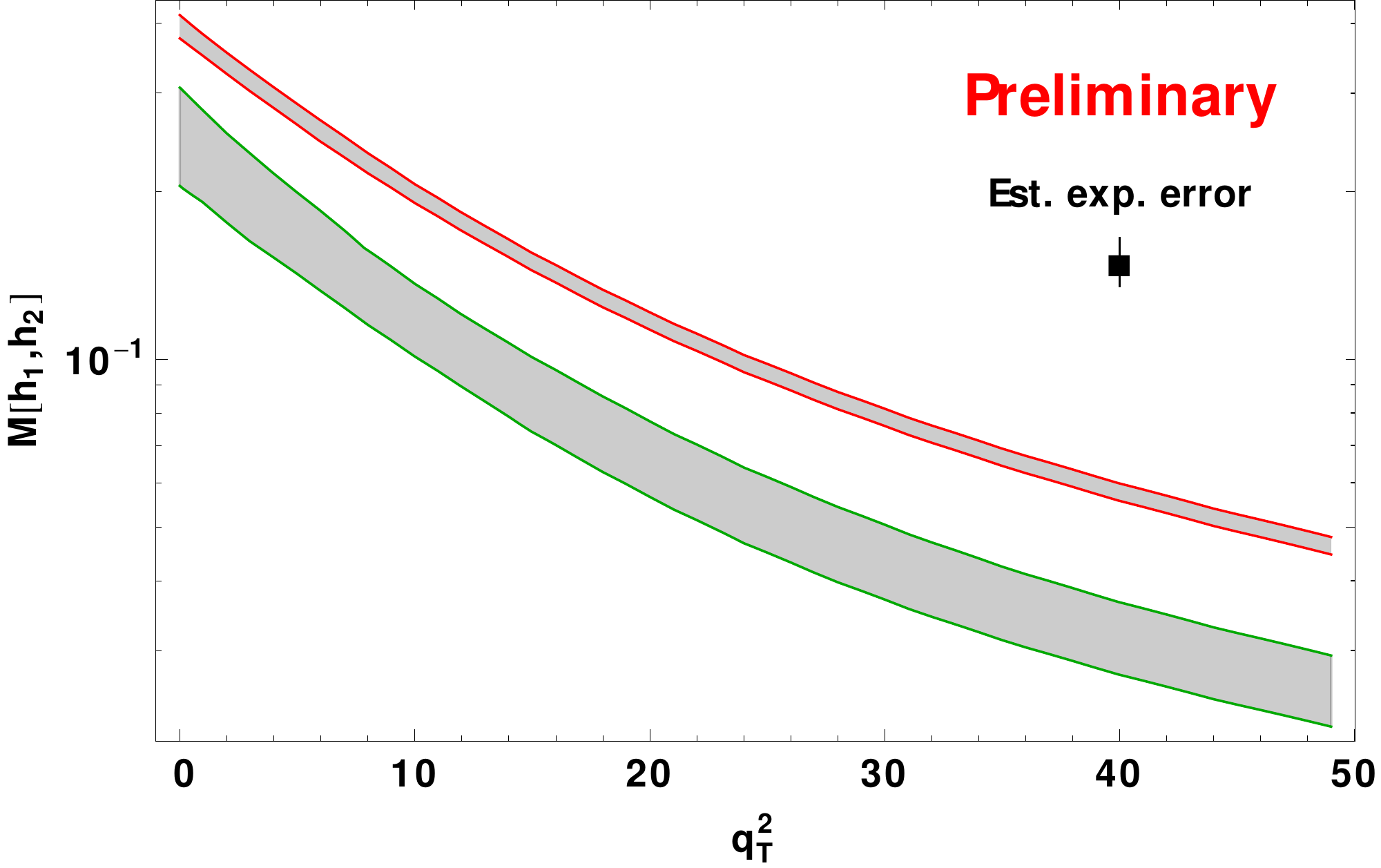}
\caption{Pion (red) vs kaon (green) multiplicities in logarithmic scale as a function of $\bm{q}^2_{\sT}$, with $z_1 = z_2 = 0.2$ and $b_{\text{max}}=1.5$ GeV$^{-1}$, $g_2 = 0.18$ (configuration C).}
\label{f:impact_fldep}
\end{figure}

In Fig.~\ref{f:impact_fldep} we compare the pion multiplicity $M[\pi^+,\pi^-]$ (in red) with the kaon one $M[K^+,K^-]$ (in green) for configuration C and $z_1 = z_2 = 0.2$.
The difference in the normalization of the two bands is due to the collinear distributions, namely the fact that it is more likely to produce pions than kaons.
The difference in the slopes, instead, is due to the different flavor combinations involved in the calculation of the cross sections.
Here two sources of flavor dependence leave their signature. The first one is connected to the intrinsic transverse momentum of quarks, namely the different Gaussian widths associated to the fragmentation processes, Eq.~\eqref{e:frag_proc}. The second one, instead, is related to the perturbative transverse momentum: evolution equations (both collinear and TMD) are flavor independent, but the initial value problem depends on the quark flavor through the initial condition $D_1^{q \to h}(z;Q=Q_0)$.
A flavor independent configuration for the TMD FFs would not result in $M[\pi^+,\pi^-]$ and $M[K^+,K^-]$ having the same slope as a function of $\bm{q}^2_{\sT}$, they would have just a different behavior from the one depicted in Fig.~\ref{f:impact_fldep} (see Ref.~\cite{Bacchetta:in_prep}). 
This means that in order to appreciate the impact of the flavor dependent TMD FFs on the multiplicities we need to disentangle the two effects. Comparisons with the future experimental data will certainly help.
Raising the $z$-value the two bands overlap (again because of the collinear distributions) but the difference in the slopes is still present.
Changing the evolution configuration does not imply substantial modifications to this result.

\subsection{Impact of kinematic dependence}
\label{ss:kinematic}

\begin{figure}
\centering
\includegraphics[width=8.5cm]{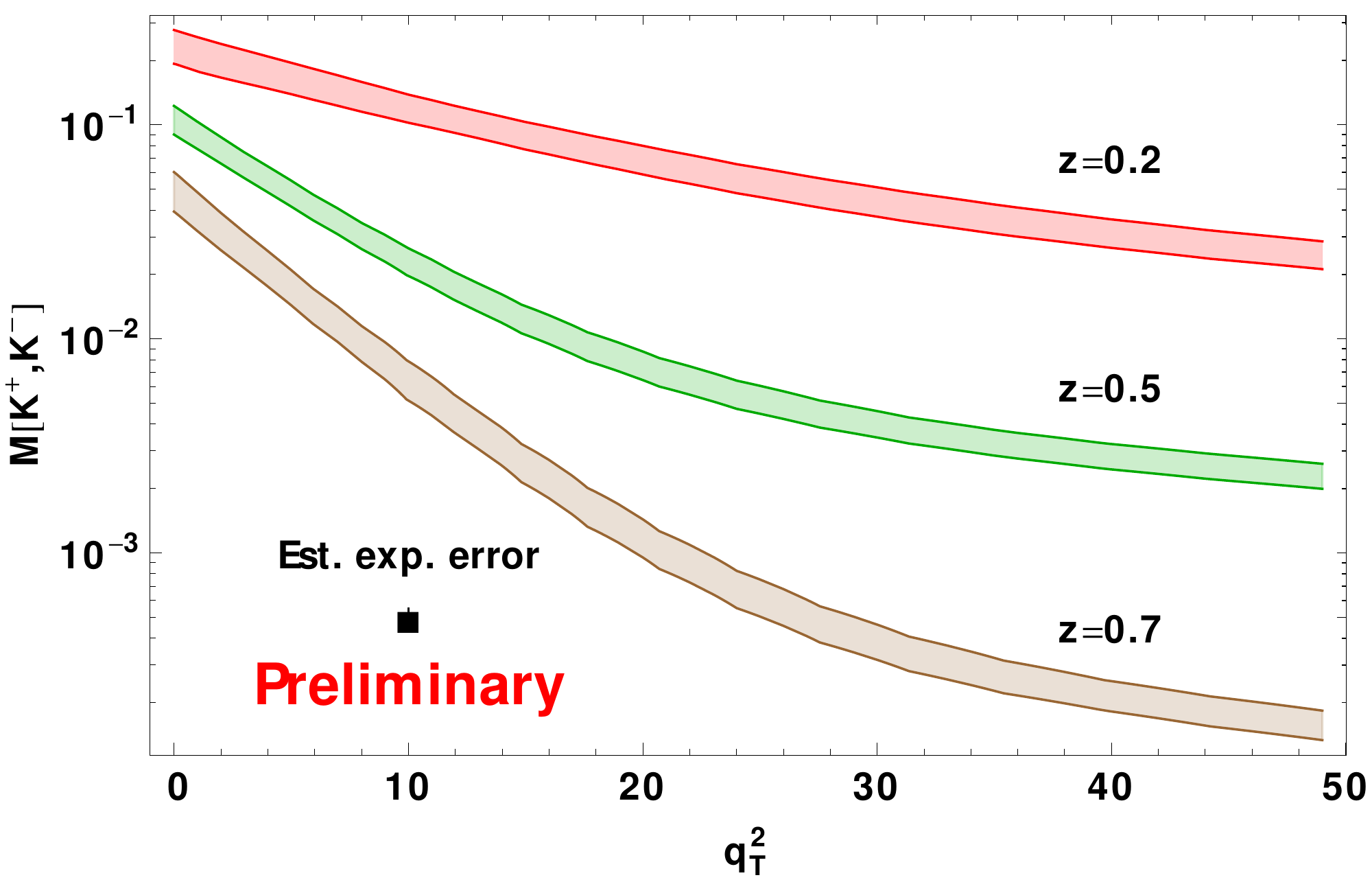}
\caption{Logarithm of kaon multiplicity $M[K^+,K^-]$ as a function of $\bm{q}_{\sT}^2$ for different $z$ values and $b_{\text{max}}=0.5$ GeV$^{-1}$, $g_2 = 0.68$ (configuration A).}
\label{f:impact_kin}
\end{figure}

In Fig.~\ref{f:impact_kin} kaon multiplicity $M[K^+,K^-]$ is plotted for configuration A and different $z$-values.
Here we can appreciate the strong impact of the $z$-dependence on the predictions. This dependence does not come only from the collinear FFs $D_1^{q \to h}(z;Q^2)$, but also from the kinematic dependence of the Gaussian widths (see Eq.~\eqref{e:z_dep}). The latter has a strong phenomenological motivation coming from the SIDIS side (see Refs.~\cite{Airapetian:2012ki,Adolph:2013stb}) and it is important to test it with $e^+e^-$ data too. 
It is also shown that with an overestimated 10\% error bar on experimental points, it will be possible to pin down the subset of replicas of kinematic parameters which reproduces experimental data best (the same holds for the flavor parameters, as evident in Fig.~\ref{f:impact_fldep}).
This ability may be reduced for pion multiplicities $M[\pi^+,\pi^-]$, because the bands would be narrower.
Changing the evolution configurations slightly affects the slope of the bands, but without major effects.

\section{Summary and conclusions}
\label{s:conclusions}

In this work we present predictions for transverse momentum dependence in pion and kaon multiplicities related to electron-positron annihilation into two hadrons.
Choosing the QCD evolution framework in Ref.~\cite{Collins:2011zzd} dressed with a model for low partonic transverse momenta inspired to Refs.~\cite{Nadolsky:2000ky,Landry:2002ix,Signori:2013mda}, we found evidence of a good sensitivity of the predictions to the non-perturbative parameters involved in the calculations.
First of all, both kaon and pion multiplicities are sensitive to variables parametrizing the high-$\bm{b}_{\sT}$ region in the evolution kernel (with some caveat concerning the $z$-values). 
Moreover, we also showed that data may be sensitive to the flavor dependence of the unpolarized TMD FFs, this resulting in different slopes of the multiplicities for pions and kaons as a function of $\bm{q}^2_{\sT}$.
Eventually, we show how the $z$-dependence in the Gaussian widths, together with the collinear $z$-dependence, affects the multiplicities.
All these features will be tested against data that will be released by experimental collaborations \belle and \babar.
We will be able to select the subset of the 200 flavor dependent TMD FFs extracted at \hermes which matches and reproduces $e^+e^-$ data best. Moreover, we will get hints on the best configuration for the non-perturbative parameters needed in the evolution kernel.
In the meanwhile, other models for low transverse momenta will be explored, together with the fixed scale evolution operator introduced in Ref.~\cite{Echevarria:2012pw}.
Updates and further developments of the current results will be presented in Ref.~\cite{Bacchetta:in_prep}.





%

\section*{Acknowledgments}
This conference proceeding is based on the talk given by AS at the ``Fourth International Workshop on Transverse Polarisation Phenomena in Hard Processes'' (Transversity 2014).
Discussions with Christine Aidala, Leonard Gamberg, Isabella Garzia, Francesca Giordano, Piet Mulders, Gunar Schnell, Ignazio Scimemi, Ted Rogers and Charlotte van Hulse are gratefully acknowledged.
The work of AS and MGE is part of the program of the Stichting voor Fundamenteel Onderzoek der Materie (FOM), which is financially supported by the Nederlandse Organisatie voor Wetenschappelijk Onderzoek (NWO). 

\bibliography{mybiblio_T14}
%

%
%
%

\end{document}